\documentclass{mn2e}

\usepackage{amssymb}
\usepackage{astrobib}
\usepackage{epsfig}
\usepackage{latexsym}
\usepackage{times}

\newcommand{\bef}{\begin{figure}}
\newcommand{\eef}{\end{figure}}
\newcommand{\beq}{\begin{equation}}
\newcommand{\eeq}{\end{equation}}
\newcommand{\ber}{\begin{eqnarray}}
\newcommand{\eer}{\end{eqnarray}}

\newcommand{\rd}{$\rho_{\rm drip}$}

\newcommand{\gcc}{\mbox{${\rm g} \, {\rm cm}^{-3}$}}

\newcommand{\msun}{\mbox{{\rm M}$_{\odot}$}}
\newcommand{\pdot}{\mbox{$\dot {\rm P}$}}

\newcommand{\lsim}{\raisebox{-0.3ex}{\mbox{$\stackrel{<}{_\sim} \,$}}}
\newcommand{\gsim}{\raisebox{-0.3ex}{\mbox{$\stackrel{>}{_\sim} \,$}}}

\title[{\it {Micro-glitch in PSR B1821-24}}]
{The micro-glitch in PSR B1821-24 : A case for a strange pulsar?}
\author[Ray Mandal et al.]
{Raka Dona Ray Mandal$^{1,3}$, 
Sushan Konar$^{2}$, 
Mira Dey $^{3}$
\and and Jishnu Dey $^{3}$\\ 
$^{1}$ Department of Physics, Rajghat Besant School, Varanasi, India \\
$^{2}$ Physics, Harish-Chandra Research Institute, Allahabad, India \\ 
$^{3}$ Ramanna Project, Department of Physics, Presidency College, Kolkata, India \\ 
e-mail : raka.raymandal@gmail.com,
sushan@hri.res.in,
mira.dey@gmail.com,
jishnu.dey@gmail.com}

\begin{document}

\date{}

\pagerange{\pageref{firstpage}--\pageref{lastpage}} \pubyear{2005}

\maketitle

\label{firstpage}

\begin{abstract}
The single  glitch observed in  PSR B1821-24, a millisecond  pulsar in
M28, is unusual on two counts.  First, the magnitude of this glitch is
at  least  an order  of  magnitude smaller  ($\Delta  \nu  / \nu  \sim
10^{-11}$) than  the smallest glitch observed to  date.  Secondly, all
other  glitching pulsars  have strong  magnetic fields  with  $B \gsim
10^{11}~G$  and are  young, whereas  PSR B1821-24  is an  old recycled
pulsar with a field strength of $2.25\times10^9~G$.  We have suggested
earlier that  some of the  recycled pulsars could actually  be strange
quark stars.   In this  work we argue  that the crustal  properties of
such a {\em strange} pulsar are just right to give rise to a glitch of
this  magnitude,  explaining  the   scarcity  of  larger  glitches  in
millisecond pulsars.
\end{abstract}


\section{introduction}
\label{sec01}

Timing irregularities seen in pulsar rotation rates are of two kinds -
{\bf  a})  timing  noise  :  continuous,  noise-like  fluctuations  in
rotation  rate; and {\bf  b}) glitches,  sudden increases  in rotation
rate, often followed by a period of relaxation towards the unperturbed
pre-glitch rotation  trend.  It is  widely believed that  these events
are caused by sudden and irregular transfer of angular momentum to the
crust   of   the   star    by   an   interior   super-fluid   rotating
faster~\cite{baym69,andr75,alpr81}.    The  result  is   a  fractional
increase  ($\delta \nu$) in  the rotational  frequency ($\nu$)  of the
pulsar,  such  that  $\sim   10^{-10}  \lsim  \delta  \nu  /\nu  \lsim
10^{-4}$~\cite{shem96,lyne00,jans06}.

In the forty  years since the pulsars were  discovered almost 300 such
glitches, large and small, have  been seen in about a hundred pulsars.
However  the March  2001  glitch of  PSR  B1821-24~\cite{cogn04} is  a
misfit both on  account of the nature of the  pulsar and the magnitude
of  the glitch. Typically  glitches are  experienced by  young pulsars
(majority have a  characteristic age $\sim 10^4 -  10^6$~yr) with high
magnetic fields ($B \sim 10^{11} - 10^{14}$~G) whereas PSR B1821-24 is
an  old (characteristic  age $\sim  3 \times  10^7$~yr) pulsar  with a
magnetic field strength  of $2.25 \times 10^9$~G.  This  is actually a
millisecond pulsar ($P  \sim 3$~ms) in the globular  cluster M28 and a
member  of  the  recycled  pulsar  population.   Moreover  the  glitch
magnitude ($\delta \nu/\nu$) is calculated to be $9.5 \times 10^{-12}$
which  is at least  an order  of magnitude  smaller than  the smallest
glitch observed so far.

The idea that  some of the pulsars could be  strange stars, instead of
neutron  stars, has  been around  for many  years~\cite{alck86}. There
have also been specific  cases like SAX J1808.4-3658~\cite{li99} which
have been proposed to be strange stars.  Contrarily it has been argued
that the magnetic field of a strange star is unlikely to decay whether
isolated  or  in  a  binary~\cite{sk00b},  whereas  according  to  the
standard  scenario of  pulsars accretion-induced  field decay  plays a
very important  role in the  evolutionary history of a  binary neutron
star~\cite{kb01}.  So there  is no definite consensus on  the case for
strange stars.  However, recycled millisecond pulsars (MSP) are stable
objects with  little or no evolution  of the magnetic field  and it is
possible for  some of them to be  strange stars.  In a  recent work we
have  argued that  the MSPs  could actually  be strange  stars  with a
limiting  value  of   the  magnetic  field~\cite{rsbkdd06}.   But  one
principal  difficulty in  explaining the  observational  properties of
pulsars  using strange  star model  has  been the  phenomenon of  {\em
glitch}.   Therefore it is  of some  interest to  note that  the MSPs,
barring B1821-24, are yet to show any significant glitch behavior even
though the  cumulative study of MSPs  is close to  $10^3$ years.  This
again goes  well with our hypothesis  that some of  the highly evolved
MSPs can be strange stars.

Therefore our aim is to establish  that a typical strange star, with a
thin  hadronic crust, can  sustain a  glitch.  And  such a  glitch, in
addition to its magnitude being consistent with that seen in B1821-24,
probably has a different origin  than the rest of glitches observed so
far.

To that intent, in Section 2  we discuss the standard theory of glitch
and note how the micro-glitch seen in B1821-24 could be of a different
nature.  In Section 3 the crustal  physics of a strange star is looked
at with particular attention  to crust-cracking and consequent glitch.
Finally in Section 4 our conclusions are presented.

\section{pulsar glitch : the energetics}
\label{sec02}

Glitches were first observed in the  Crab and the Vela and they mainly
occur  in  younger  pulsars.   While  the classical  glitch  has  been
regarded as a  sudden increase in rotational frequency  followed by an
exponential  recovery  of a  large  fraction,  it  is not  typical  of
glitches  in most  pulsars.  The  dominant  effect is  an increase  in
frequency with  very little recovery~\cite{shem96,lyne00,kraw03}.  The
standard theory describes a glitch  as an event in which a significant
number  of  vortices are  suddenly  unpinned  from  the crust  nuclei,
angular momentum  is transferred  to the crust,  and the  vortices are
eventually  re-pinned.  Wherever  recoveries are  seen  they typically
have a relaxation  time of the order of days to  months which has been
explained by invoking an interaction of the crust with the super-fluid
component  in the  interior~\cite{baym69,saul89}.  In  particular, the
so-called  {\em  vortex creep}  model  is  known  to provide  adequate
description                of                the                glitch
relaxation~\cite{alpr81,alpr84,alpr93,alpr96}.   The  actual mechanism
that triggers the glitch remains unspecified in almost all models.

Though  a large  number of  pulsars exhibit  glitches,  certain points
about  the glitching pulsars  need to  be noted.   Fig.\ref{fig01} and
fig.\ref{fig02}  show histograms  of the  characteristic ages  and the
inferred magnetic  (dipolar) field strengths of  the glitching pulsars
(for a complete list of known glitching pulsars see table-I of Melatos
et al.~\citeyear{mela08}). It is obvious that these pulsars are mostly
relatively young. Only a few  have spin-down ages similar to B1821-24,
but none of them are of  the recycled variety.  And no pulsar, barring
B1821-24,  with  a magnetic  field  smaller  than  $10^{11}$G shows  a
glitch.  

Moreover if  we look  into the  energy budget  of  the glitching
pulsars an interesting fact  emerges. The rotational kinetic energy of
a pulsar can be approximately given by,
\ber
E_{\rm rot} \sim I \nu^2 \,, \nonumber \eer
where  $I$ is  the moment  of inertia  of the  star and  $\nu$  is the
observed spin-period,  apart from a  numerical factor of  $\sim10$. Of
course, the  observed $\nu$  refers to that  of the crust.   Since the
super-fluid  component  rotates  faster  than  the  crust  the  actual
rotational   energy  would   be   somewhat  larger   than  the   above
estimate. But  it would be of the  same order and we  shall ignore the
difference for  the present discussion.   A glitch in  the spin-period
implies a change in the rotational energy, $\Delta E$, given by
\beq
\Delta E \,  = \, \delta \, (I \nu^2)  \sim I \nu^2 \left(\frac{\delta
\nu}{\nu}\right)\,  \sim \left(\frac{\delta \nu}{\nu}\right)  \ E_{\rm
rot},
\eeq
where $\delta \nu$ is the  magnitude of the glitch. Evidently, $\Delta
E$  is  the energy  scale  associated  with a  glitch  and  it can  be
estimated for a  glitching pulsar using the observed  value of $\delta
\nu$.   In fig.\ref{fig03}  we plot  $\Delta E$  assuming  the stellar
moment of inertia to be  $\sim 10^{45}$gm.cm$^2$ for all the glitching
pulsars (we shall again ignore  the small differences in $I$), showing
that  the  energy  scale associated  with   glitches  has a  range  of
$10^{36}-10^{44}$erg.  

At once the  fact that B1821-24 is different  from the other glitching
pulsars becomes  evident. Even  though $\Delta E$  ($\sim 10^{40}$erg)
itself falls quite  within the above range it is  extremely high for a
glitch of  such small magnitude. This  is simply due to  the fact that
the rotational kinetic  energy of an MSP is  very large.  require much
larger energy  change in an MSP than  in a normal pulsar.   Now if for
some  reason  such  large  energy  scales,  required  for  large  size
glitches, are  not available to MSPs  it would explain  the absence of
such glitches in them.  We show that exactly that would happen if MSPs
are assumed to be strange stars.

\bef
\epsfig{file=fig01.ps,width=165pt,angle=-90}
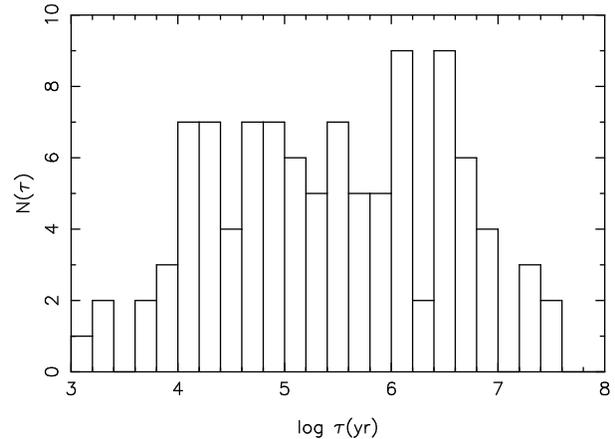
\caption[]{Histogram showing  the distribution of  characteristic ages
($\tau$)  of 104  pulsars  reported to  experience  one or  more
glitches.  The data  has been taken from Melatos  et al.(2008) and the
ATNF on-line catalog.}
\label{fig01} 
\eef

\bef
\epsfig{file=fig02.ps,width=165pt,angle=-90}
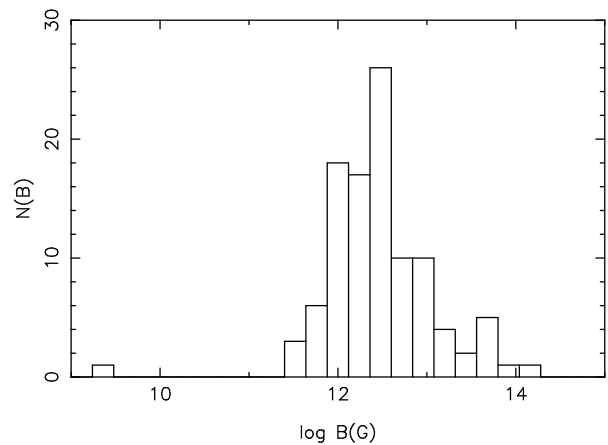
\caption[]{Histogram showing the distribution of magnetic fields ($B$)
of 104 glitching pulsars mentioned in Fig.\ref{fig01}.}
\label{fig02} 
\eef

\bef
\epsfig{file=fig03.ps,width=165pt,angle=-90}
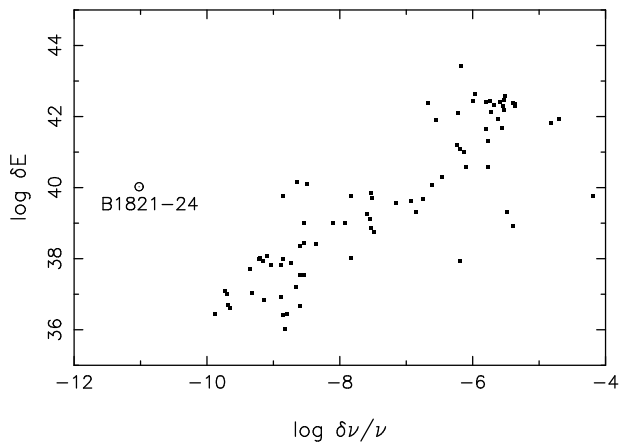
\caption[]{Change in the rotational energy ($\delta E$) vs. the glitch
magnitude  ($\delta \nu/\nu$) assuming  the moment  of inertia  of the
compact object  to be $\sim  10^{45}$gm.cm$^2$.  The maximum  value of
$\delta  \nu/\nu$  has been  used  for  pulsars experiencing  multiple
glitches.   The point  corresponding  to PSR  B1821-24 clearly  stands
out. The  data has  been taken from  Melatos et al.(2008)  (only those
where a  value of $\delta \nu/\nu$  is available) and  the ATNF on-line
catalog.} 
\label{fig03} 
\eef

In  this context,  it would  be helpful  to recapitulate  the standard
wisdom regarding  the MSPs.  A pulsar  is understood to  be a strongly
magnetized rotating neutron star.   The measured spin-period ($P$) and
the  estimated dipolar  component of  the magnetic  field  ($B \propto
\sqrt{P\pdot}$,  where  $\pdot$  is  the  period  derivative)  broadly
classify the  pulsars in  two categories -  {\bf a})  isolated pulsars
with rotation periods usually above 1s and very strong magnetic fields
($10^{11} - 10^{14}$~G); {\bf b}) binary/millisecond pulsars with much
shorter  rotation  periods  and  considerably weaker  magnetic  fields
($10^{8} - 10^{10}$~G).

Observations suggest a connection  between the second group with their
being processed  in binary systems, prompting  theoretical modeling of
accretion-induced        reduction        of       the        magnetic
field~\cite{kb97,kb99a,kb99b,cumm01,ck02,kc04,payn04,payn07}.        In
fact, the discovery of SAX  J1808.4-3658, a 2.49~ms X-ray pulsar, with
an  estimated dipole  field  strength  of $\sim  10^8$~G  is a  direct
pointer  to  the  connection   between  low-mass  X-ray  binaries  and
MSPs~\cite{wijn98,chak98}.   It is understood  that this  object would
emerge as a  typical millisecond radio pulsar once  the mass accretion
in the system stops, vindicating present theoretical expectations. For
obvious reasons the members of the  second group are also known as the
recycled pulsars.

It should  be noted that in our discussion,  we have not included
the 16ms pulsar  J0537-6910~\cite{midl06}, observed to experience many
large  glitches, because  this is  a  young pulsar  associated with  a
supernova  remnant  and  is  not  likely to  belong  to  the  recycled
population of typical MSPs.  

\section{strange star : crust-cracking and glitch}
\label{sec03}

It is understood that a strange star can form via the deconfinement of
the  nuclear matter  in  an accreting  neutron star~\cite{webr99}.   A
1.4\msun~neutron star  requires to accrete $\sim  0.5$\msun~or more to
attain   the  deconfinement   density   ($\rho_{\rm  deconfine}   \sim
8\rho_{\rm nuc}$) at the  center~\cite{chen96}.  This is possible only
if the  companion is a low-mass  star, as the amount  of accretion has
been   shown   to    range   from   0.1\msun~to   1\msun~in   low-mass
binaries~\cite{bhat91,verb93,bitz95}.   On the  other hand,  a neutron
star  in a low-mass  binary is  also known  to produce  the ubiquitous
MSPs~\cite{db02,wijn98}.  Evidently, the MSPs are prime candidates for
the strange stars.

The interior of  a strange star supposedly consists  of a u-d-s plasma
with a small admixture of  electrons, to make it charge-neutral, where
each  particle species  is Fermi-degenerate.   Strange  star structure
modeled with a  realistic equation of state has  been developed by Dey
et al.~\citeyear{deym98}.  Furthermore, the u-d-s plasma could be in a
superconducting state  where the rotation  and the magnetic  field are
supported  by formation of  vortex bundles~\cite{rsbkdd06}.   Here, we
consider  such a  rotating strange  star  which also  supports a  thin
hadronic crust.   The maximum  density at the  bottom of  the hadronic
crust  of a strange  star is  that of  the neutron  drip (\rd  $\sim 4
\times 10^{11}$~\gcc)~\cite{glen95}.  This crust is separated from the
interior  by an  electrostatic gap  of  few Fermi  which prevents  the
nuclear  matter from  further  conversion.  In  our  model, a  typical
1.4\msun~star would  have a  $\sim 100$~cm thick  crust of  mass $\sim
10^{-5}$\msun.

The primary region of interest  in the case of any timing irregularity
of a pulsar  is the crust (whatever may be the  nature of the coupling
of  the crust  to  the interior).  Therefore  is it  of importance  to
understand the  nature of the  crust. 
For  the sake  of convenience  we  shall assume  that the  crust of  a
rotating strange  star consists of  cold-catalyzed matter (equilibrium
nuclide  corresponding to  a particular  density at  zero temperature)
where      the     density      varies     from      $7.86$\gcc     to
$4\times10^{11}$\gcc~\cite{baym71}.  The melting temperature of such a
crust ranges from  $\sim 10^8 - 10^9$K and  therefore it is reasonable
to assume the crust to be in a crystallized state~\cite{gudm83}.
The lattice spacing
of  this crystal is  $\sim 10^{-11}-10^{-10}$cm  which is  much larger
than the nuclear  size implying that the crust  behaves like a Coulomb
crystal.

The shear stress of this crystal for a particular density is  given by
$\sigma_{\rm   shear}  \sim   \mu  \theta$   where  $\theta$   is  the
dimensionless strain~\cite{rudr91a,rudr91b}. The shear modulus, $\mu$,
is given by :
\ber
\mu \simeq\frac{0.3 (Ze)^2}{a^4}, \; \; \;  {\rm with} \; \; \; a \sim
\left(\frac{\rho}{m_a A}\right)^{-1/3} \nonumber
\eer
where $(Z,A)$ correspond to  the equilibrium nuclide at density $\rho$
and  $m_a$ is  the atomic  mass unit~\cite{ashc}.   The  maximum shear
stress which a  crustal lattice can support is  uncertain.  It depends
upon the  number of lattice dislocations, and  their location, pinning
and mobility.  In the case of  a neutron star, it has been argued that
the strain  angle $\theta$ should  typically be between  $10^{-5}$ and
$10^{-3}$   as   the   crustal   lattice  behaves   like   an   alkali
metal~\cite{smol70a,smol70b}.  The  lower values are  more probable as
the crust is  likely to have many defect  points (impurity ions etc.).
See   Horowitz    \&   Berry~\citeyear{horo09b}   and    Horowitz   \&
Kadau~\citeyear{horo09a} for recent work  on the strength of the crust
of a neutron  star.  The hadronic crust of a  strange star is expected
to be  very similar to  the outer crust  of a neutron star.   We shall
therefore assume the above values to hold for our discussion too.

The shear stress at the bottom of the hadronic crust of a strange star
is then given by,
\beq
\sigma_{\rm   shear}    \simeq   10^{29}   \times    \left(10^{-4}   -
10^{-5}\right) \; {\rm dyne \ cm}^{-2},
\eeq
where we  have assumed  a range of  $10^{-4} - 10^{-5}$  for $\theta$.
However  it should  be noted  that we  are considering  the case  of a
strange star  that has formed  via the deconfinement conversion  of an
accreting neutron  star.  The  crust of an  accreting neutron  star is
unlikely  to be  composed of  cold-catalyzed matter.   In  addition to
having  a  large  number of  defects  (hence  a  very small  value  of
$\theta$), the dominant nuclei at a given density could also have much
lower  values  of  $Z$  and  $A$  as  these  are  generated  by  local
shell-burning  processes  induced  by  accretion~\cite{brwn98,scha01}.
This might  actually work  to reduce the  above value of  shear stress
further.

Therefore, the maximum energy associated  with the shear stress of the
entire hadronic crust of a strange star is
\beq
E_{\rm  shear} \sim \sigma_{\rm  shear} V_{\rm  crust} \sim  10^{40} -
              10^{41} \ {\rm erg},
\label{eq_Es}
\eeq
where $V_{\rm crust}$ is the volume  of the crust that has a thickness
of $\sim$100~m  on a  star of total  radius $\sim$10~km. It  should be
noted that this  estimate is of the maximum  shear energy available to
the crust, as  the calculation for the shear stress  has been based on
the bottom layer  of the crust.  In general the  top layers have lower
shear stress and the total shear energy could be smaller by one or two
orders of magnitude.  Still, it is clear that if the energetics of any
process is such that the stress  on the crust becomes very much larger
than $\sigma_{\rm  shear}$ the  crust would give  in to  plastic flow.
Evidently,   there    could   be   no   {\em   quake}    in   such   a
situation. Therefore  it could be  concluded that while  energy scales
similar  to that  estimated in  Eq.(\ref{eq_Es}) would  quite possibly
give  rise  to micro-glitches  (similar  to  that  seen in  B1821-24),
energies much  larger than this would  most likely not  be observed as
star-quakes resulting in glitches.

\section{Discussions and Conclusions}
\label{sec04}

\bef
\epsfig{file=fig04.ps,width=165pt,angle=-90}
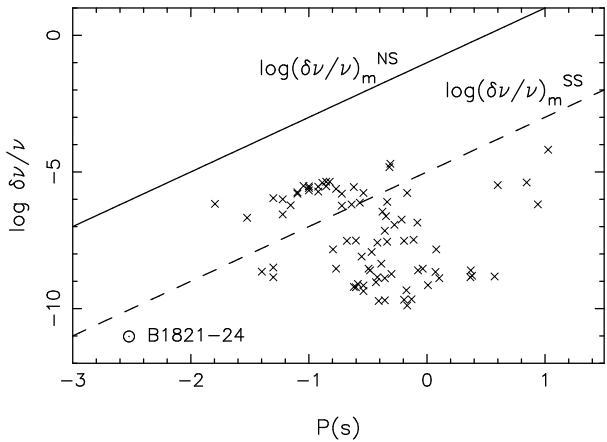
\caption[]{The  magnitude of  glitch, $\delta  \nu/\nu$,  (maximum for
pulsars with multiple glitches) vs. the spin period ($P$) of glitching
pulsars.  The  straight lines  denote the theoretical  predictions for
maximum $\delta  \nu/\nu$ vs. $P$  for neutron stars (NS)  and strange
stars (SS). As expected B1821-24 falls below the SS line and there are
{\em no} glitching  pulsars above the line corresponding  to a neutron
star. The data is the same as in Fig.\ref{fig03} above.}
\label{fig04} 
\eef

It is essential  to realize that a very  different physical phenomenon
would give rise to a glitch in  a strange star than in a neutron star.
The  glitch  in  a  neutron  star  is  essentially  a  result  of  the
weak-coupling between  the neutron  super-fluid with the  crust, which
rotates  slower than the  neutral super-fluid.   But such  a situation
does  not  arise  in  a   strange  star.   Even  if  the  quarks  form
super-condensates in the  interior they would be coupled  to the crust
via electromagnetic  interaction because the quarks  are charged.  The
timescale  for  electromagnetic coupling  is  so  small  that for  all
practical purposes the core  and the crust should co-rotate.  However,
it has been understood early on  that some glitches could very well be
signatures  of star-quake  phenomena~\cite{baym69}.   We believe  that
could be the case for B1821-24 too.

It needs to be noted that there has been recent work to model directly
the dynamics of crust cracking and vortex avalanches.  Extended timing
observations of  J0537-6910 indicate that  the time interval  from one
glitch to  the next glitch is  strongly correlated to  the amplitude of
the first  glitch, a pattern  that is quite  similar to that  of large
quakes  within the  crust  of our  planet~\cite{midl06}.   It is  also
likely  that  at least  some  glitches, like  those  in  the Crab  and
B0540-69 may  be crust-quakes, where the  equilibrium configuration for
the  solid crust  departs from  its geometrical  configuration  as the
pulsar  spins down  until  eventually the  crust  cracks and  settles.
Recent analysis of glitch data  also suggest that glitches result from
scale-invariant avalanches, which are consistent with a self-organized
critical   system~\cite{mela07,mela08}.   An   early  model   of  such
self-organized    criticality    was    developed   by    Morley    \&
Schmidt~\citeyear{morl96} by assuming the crust of the neutron star to
consist of a number of plates. These plates could be strained due to a
deviation from  the equilibrium configuration  of the crust  and their
relaxation  at  the  point  of  maximal stress  would  then  induce  a
crust-cracking event.   Recently this  model has been  investigated in
detail   by   Warszawski   \&  Melatos~\citeyear{wars08}   and   their
theoretical   expectations  match  well   with  the   observations  of
Middleditch et al.~\citeyear{midl06}.

In  this context,  it  would be  interesting  to look  at the  crustal
strength of a neutron star itself,  in the spirit of the discussion in
section-\ref{sec03}.   The crust  of  a neutron  star  is about  1$km$
thick,    has   a    density    range   of    $7.86$\gcc   to    $\sim
10^{14}$\gcc~\cite{baym71} and  behaves like a coulomb  crystal at all
times       (barring      few       years       immediately      after
birth)~\cite{gudm83,page98}.  Following the line of reasoning above we
find that the  maximum energy associated with the  shear stress of the
entire hadronic crust of a neutron star is
\beq
E_{\rm  shear} \sim  10^{44} - 10^{45} \ {\rm erg},
\label{eq_En}
\eeq
where  we  have again  assumed  a range  of  $10^{-4}  - 10^{-5}$  for
$\theta$.   From this  we can  estimate the  maximum value  of $\delta
\nu/\nu$ for  a pulsar of  a given spin-period.  $\delta  \nu/\nu$ vs.
$P$ has  been plotted  for all known  pulsars in  fig.\ref{fig04}. For
comparison we plot  the theoretical curves for the  maximum of $\delta
\nu/\nu$ corresponding  to a neutron star  as well as  a strange star.
Not surprisingly, the magnitude of  all the observed glitches are well
below  the  maximum  value  calculated  for a  neutron  star.   It  is
interesting to  note that  even though there  are a number  of pulsars
residing between  the two  lines, B1821-24 is  not one of  them.  Even
though  this  is  a  very  happy  situation,  a  word  of  caution  is
necessary. The crust  in a neutron star is far  more complex than that
in a strange  star. To begin with, one needs  to consider the presence
of the neutron  superfluid beyond the neutron drip  density ($4 \times
10^{11}$\gcc).   Then, in  the  deeper  layers of  the  crust, as  the
nuclear  density  is  approached,   the  nuclei  themselves  may  have
non-spherical  shapes~\cite{lorn93}  giving  rise  to  very  different
structural  properties.  Therefore,  the above  estimates  for $\delta
\nu/\nu$ maximum may not be very reliable.

However,  it can  be seen  that it  is possible  to conceive  of other
causes  giving  rise to  crust-cracking  stresses  in  a strange  star
resulting  in generalized  star-quakes, even  though there  can  be no
vortex  dynamics  associated with  the  strange  star  crust.  Such  a
mechanism for strange star quake has already been discussed by Peng \&
Xu~\citeyear{peng08} in  the context of the  slow-glitches observed in
PSR B1822-09.  We  too believe that some kind  of crust-cracking event
resulted in the star-quake which was observed as the March 2001 glitch
in PSR B1821-24.

In  fact, observations  of  a larger  number  of events  like that  in
B1821-24  is required for  a clearer  picture of  the MSP  glitches to
emerge. Though accumulated glitch  data currently indicates a relation
between the glitch rate ($\lambda$)  and the spin-down age ($\tau$) no
definite relation is available yet. However, it is understood that for
older pulsars $\lambda$ is expected to be much smaller than $0.25$ per
year~\cite{mela08}.  Cognard  \& Backer~\citeyear{cogn04} estimate the
cumulative number of years of MSP timing observations up to 2001 to be
around 500  (hence a much  larger value at  the present date)  for all
objects.   Thus, a maximum of  about 125  glitches could  have been
observed  in  MSPs.  So,  the  current  data  is consistent  with  the
expectation that the older pulsar glitch infrequently. Only if a large
number of MSP glitches are observed  in future this idea would have to
be re-evaluated.   However, neither can  the possibility that  some of
the MSPs are  strange stars be ruled out as they  too are not expected
to  experience frequent  glitches except  as some  crust-quake events.
Moreover,  an assumption  of strange  star automatically  explains the
scarcity of larger glitches in MSPs.

\bef
\epsfig{file=fig05.ps,width=165pt,angle=-90}
\caption[]
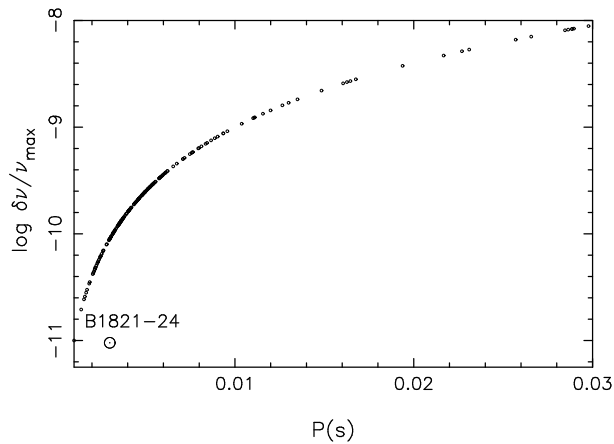
{The maximum value  of $\delta \nu/\nu$ predicted for  the known MSPs,
assuming them  to be  strange stars.  PSR  B1821-24 is well  below the
predicted value. The MSP periods have been taken from the ATNF on-line
catalog  and Paulo C.   Freire's catalog  of globular  cluster pulsars
({\em http://www.naic.edu/~pfreire/GCpsr.html}).}
\label{fig05} 
\eef

It  is evident from  the discussions  in section-\ref{sec03}  that the
energy-scales  associated   with  the  micro-glitch   of  B1821-24  is
equivalent  to   the  maximum  shear  stress  of   the  entire  crust.
Therefore, if B1821-24 is indeed a  strange star then we do not expect
to   see  a   glitch  of   larger  magnitude   in  this   pulsar.   In
fig.\ref{fig05} the maximum  values of $\delta \nu/\nu$, corresponding
to the maximum shear stress in  a strange star crust, has been plotted
against     the    spin-periods     of    all     known    millisecond
pulsars~\cite{lorm08}. Any future observation of a glitch with $\delta
\nu/\nu$ larger  than the  maximum calculated here  should be  a clear
indication against the hypothesis of MSPs being strange stars.

To conclude we note that the  absence of glitches in MSPs could be due
to the simple  fact that older pulsars glitch  infrequently or because
they are strange  stars. But we need to observe  many more glitches in
MSPs to settle this question properly.

\section{acknowledgments}

This work has  made extensive use of the data  from ATNF on-line pulsar
catalog~\cite{manc05}            available           at           {\em
http://www.atnf.csiro.au/research/pulsar/psrcat/}.    Also,  we  would
like to acknowledge Paulo C. Freire for his on-line catalog of globular
cluster pulsars.

We would  like to thank John  C. Miller for  originally attracting our
attention to  this problem. RDRM  thanks Asit Banerjee,  B.~P.  Mandal
and  T.~V.  Ramakrishnan for  illuminating  discussions.  We are  also
grateful to  the anonymous referee  for making some  valuable comments
which have helped improve the paper substantially.

\bibliography{mnrasmnemonic,refs}

\bibliographystyle{mnras}

\bsp

\label{lastpage}
\end{document}